\documentclass[aps,pre,singlecolumn,showpacs]{revtex4-1}  
\usepackage[dvipsnames]{xcolor}
\usepackage{graphicx}  
\usepackage{dcolumn}   
\usepackage{bm}       
\usepackage{amssymb}  
\usepackage[normalem]{ulem}
\usepackage{amsmath}
\usepackage{dsfont}
\usepackage{xcolor}
\usepackage{float}
\usepackage{enumerate}
\usepackage{braket}
\begin{document}
\title{Quantum correlations and their significance in quantum teleportation under $\cal{PT}$-symmetric operation}
\author{ J. Ramya Parkavi$^1$,  R. Muthuganesan$^2$,  V. K. Chandrasekar$^1$, and M. Lakshmanan$^3$}
\address{$^1$Department of Physics, Centre for Nonlinear Science \& Engineering, School of Electrical \& Electronics Engineering, SASTRA Deemed University, Thanjavur -613 401, Tamilnadu, India\\
$^2$Department of Physics, Faculty of Nuclear Sciences and Physical Engineering, \\Czech Technical University in Prague,
B\u rehov\'a 7, 115 19 Praha 1-Star\'e M\u{e}sto, Czech Republic	\\
$^3$Centre for Nonlinear Dynamics, School of Physics, Bharathidasan University, Tiruchirappalli 620 024, Tamilnadu, India.}
 \begin{abstract} 
In this article, we exploit the different notions of quantumness measure to  understand the properties of the Heisenberg XY model with and without $\cal{PT}$-symmetric operation. In the absence of $\cal{PT}$-symmetry, we study the significance of different measures, namely entanglement and measurement-induced nonlocality (MIN), in the detection of the quantumness of the Heisenberg XY model. It is observed that the quantum correlations and teleportation fidelity monotonically decreases with respect to temperature. Furthermore, the intervention of $\cal{PT}$-symmetric operation enhances the strengths of quantum correlation. In addition, we highlight the role of the system's parameters and  $\cal{PT}$-symmetric operation on the teleportation of a quantum state. Our results also emphasize that after the addition of $\cal{PT}$-symmetric operation, the considered physical model remains a versatile resource to achieve successful teleportation of the quantum state.

\end{abstract} 
\maketitle
\section{Introduction}
Characterizing quantum correlation in composite systems via local measurements/operations is one of the fundamental areas of research in quantum information theory. One of the most fundamental quantum correlation measures is entanglement, which was introduced by Schr$\ddot{o}$dinger \cite{Schrodinger1935} due to the paradox demonstrated by Einstein, Podolsky, and Rosen (EPR) \cite{Einstein1935}.  It is argued that due to the entangled nature of systems, the measurement on one of the subsystems will affect the outcome of measurements performed in the other subsystem. This feature obviously contradicts local realism, i.e., the quantum states of spatially separated non–interacting particles are interdependent. The contradiction arising in the above argument leads to the question of the completeness of quantum mechanics which was rooted in their insistence on local realism. Thus, entanglement is viewed as ``spooky action'' at a distance exhibited in the composite quantum systems by virtue of the superposition principle. In an entangled composite quantum system, the individual subsystems are strongly linked to each other even if they are far apart and no longer interact with each other. The whole composite quantum system can be well described as a definite pure state while for the state of its subsystems, such a description is not possible.

Subsequently, assuming local realism of EPR, Bell attempted for the completeness of quantum mechanics using the hidden variable theory \cite{Bell1964,Bell1966}. One of the main manifestations of nonlocal attributes of many body quantum states is entanglement which has no classical analog. In quantum information theory, entanglement is used as a resource for different applications: for instance, in quantum cryptography, or teleportation schemes, or quantum computation, where it is used much like energy or information that are considered as resources in classical processing. While the entanglement between bipartite (two particles) states has been extensively investigated over many decades, the entanglement of multipartite (many particles) states are increasingly difficult for investigation. Even within the domain of bipartite states, it is now fairly realized that entanglement is not the complete manifestation of non-classical correlation between the particles.

In order to capture the complete spectrum of nonlocality, a new measure of quantum correlation from the perspective of measurement called quantum discord  \cite{Ollivier2001} has been introduced. The operational meaning of this quantity is the minimal loss of information of a state due to local quantum measurements. Computation of discord involves complex optimization \cite{Girolami2011} and is also shown to be a nondeterministic polynomial problem \cite{Huang2014}. For easy computation, till date, various quantum correlations have been identified based on the metric in state space \cite{Girolami2012,Dakic2010,Luo2011,Spehner2017,Roga2016}. Measures based on the metric are experimentally realizable \cite{Jin2012, Passante2012, Girolami2012}. One of the notable correlation measures beyond entanglement is the measurement induced nonlocality (MIN) \cite{Luo2011}. MIN characterizes the nonlocality from a measurement perspective, which is different from the conventionally mentioned quantum nonlocality related to entanglement or the violation of Bell's inequalities. This quantity can be considered as an indicator of the global effect caused by locally invariant measurements. Up to date, several other versions of MIN were suggested too \cite{HFan2015,MuthuPLA,XiMIN,HuMIN,HFanReview}.

\par On the other side, non-Hermitian systems with parity-time ($\cal{PT}$) symmetry proposed by Bender and Boettcher have attracted growing interest over the past two decades. These systems support two distinct phases, namely unbroken phase and broken phase. In these systems, the eigenvalues are real in $\cal{PT}$ unbroken regime and the eigenvalues are complex in the spontaneously broken $\cal{PT}$ regime \cite{ptbase1,ptbase2, ptbase3}. As a result, the dynamics is drastically different in these two phases, and a dynamical criticality occurs at the boundary between the two regimes called exceptional points. The  exploitation of quantum mechanical phenomena around these exceptional points is a formidable task in the domain of quantum information theory. Recently, their realization in different $\cal{PT}$- symmetric systems \cite{ijtp, ourpaper, bellPT} and then their applications were demonstrated.  For instance, the application of local non-Hermitian detuning on entangled bipartite state significantly reduces the uncertainty of quantum measurements \cite{entropynonherm}. In particular, the information flow between a $\cal{PT}$-symmetric non-Hermitian system and its environment has been studied theoretically as well as experimentally and it has been shown that the complete retrieval of information  is possible in the unbroken $\cal{PT}$-symmetric regime\cite{infflowtheory, infflowexp}. In Ref. \cite{rocatti}, F. Roccati et al. studied the dynamics of quantum correlations in a $\cal{PT}$-symmetric gain-loss system which exhibits stationary quantum correlations beyond entanglement. They also pointed out the usefulness of such stationary quantum correlations in information encoding \cite{infencod}, remote-state preparation \cite{remote}, quantum metrology and sensing \cite{quanmetrology}, etc. To an extent, the authors of \cite{highpt} experimentally demonstrated the quantum information dynamics through entropy in a high dimensional $\cal{PT}$-symmetric system with its exceptional point degeneracies. 

\par Compared to the above systems, the Heisenberg spin chain model is noteworthy which accounts for the enormous phenomena in magnetic systems which can be implemented in the exploitation of quantum/magnetic phase transitions, quantum correlations and teleportation. For example, teleportation which depends on the entangled states was demonstrated with the effects of dipolar interaction between two spin-$1/2$ particles and it was shown that even without the violation of Bell's inequality all the entangled states are useful for teleportation \cite{dipolar}. This is in contrast to the earlier studies which showed that any two spin-1/2 state which violates the Bell-CHSH inequality proposed by Clauser, Horne, Shimony and Holt is useful for teleportation \cite{fidelbellreferee1, fidelbellreferee2}. The authors of Ref. \cite{DZ} also highlighted that a minimal entanglement of the thermal state is needed to realize the entanglement teleportation in a two-qubit Heisenberg chain model with Dzyaloshinski-Moriya interaction. 
\par However, to our knowledge, fewer research works related to the quantum correlations and their significance in quantum teleportation have been investigated, especially under the action of $\cal{PT}$-symmetric operation. Two questions are putforth here: (i) How does the $\cal{PT}$-symmetric operation affect the existence of quantum correlations in a Heisenberg XY spin system? and  (ii) Whether the $\cal{PT}$-symmetric nature enhances or diminishes the teleportation process in the system?  To answer these queries, in our work, we mainly investigate the co-evolution of quantum correlations and quantum teleportation of the Heisenberg  XY spin model under the non-unitary time evolution. We have measured the thermal as well as temporal quantum correlations via four different measures, namely Bell nonlocality, concurrence, measurement induced non-locality, and trace distance based measurement induced nonlocality. In addition to these, we try to exploit the teleportation process even in the absence of entanglement. 

\par The article is structured as follows: First, we introduce the different measures of quantum correlations in Sec. II. The considered Heisenberg XY model is described in Sec. III. We display the results of thermal quantum correlations and thermal teleportation for the considered model in Sec. IV.  The role of $\cal{PT}$-symmetric operation is explored in the model in Sec. V. In Sec. V A, the quantum correlation measures and teleportation under $\cal{PT}$-symmetric time evolution will be discussed. Finally, we summarize the key findings of this paper in Sec. VI.
\section{Preliminaries}
In this section, we introduce the quantum correlation quantifiers of bipartite states. For this purpose, we consider a bipartite system represented by its density matrix $\rho$ shared between the subsystems $a$ and $b$.

\textbf{Entanglement:}
The amount of bipartite entanglement between the subsystems is measured by the quantity concurrence and it is defined as \cite{concurform}
\begin{align}
\mathcal{C}(\rho)=2~ \text{max}~\left\{0,~ \lambda_1- \lambda_2-\lambda_3-\lambda_4 \right\},
\end{align}
where $\lambda_i$'s are the square root of the eigenvalues of the matrix $\Lambda=\sqrt{\rho}\tilde{\rho} \sqrt{\rho}$ and arranged in decreasing order. Here $\tilde{\rho}=(\sigma_y \otimes \sigma_y)\rho^*(\sigma_y \otimes \sigma_y)$ is the spin flipped density operator and the symbol $*$ denotes the usual complex conjugate in the computational basis. The well-known fact is that concurrence takes the minimum and maximum values from $0$ to $1$  which correspond to the separable (unentangled) and maximally entangled states, respectively.

\textbf{Maximal Bell function:}

In general, for any qubit-qubit system, Bell inequality is a well-known technique to demonstrate the presence of nonlocality of a state. It is also useful in the quantification of quantum correlation in physical systems. The nonlocality of a quantum state can be detected by the violation of the Bell-CHSH inequality, which is given by \cite{Clauser}
\begin{equation}
|\langle \mathcal{B}_{{\text{CHSH}}_{\rho}} \rangle| \leq 2,
\end{equation}
where $|\langle \mathcal{B}_{{\text{CHSH}}_{\rho}} \rangle|=\text{Tr}(\rho ~\mathcal{B}_{\text{CHSH}})$ and $\mathcal{B}_{\text{CHSH}}$ is the Bell operator associated with the quantum CHSH inequality. To quantify the quantum correlation between the subsystems via Bell inequality, the maximal Bell function is defined as \cite{Horodecki, Brunner}
\begin{equation}
\mathcal{B}_{\text{max}}=2 \sqrt{M(\rho)},
\end{equation}
where $M(\rho)=\text{max}_{i < j}(R_i+R_j)$ with $R_{i}(i=1,2,3)$ being the eigenvalues of the correlation matrix $R^{t}R$. The matrix elements of $R$ are defined as $r_{ij}=\text{Tr}(\rho (\sigma_i \otimes \sigma_j))$. The function $\mathcal{B}(\rho)>2$ indicates the violation of Bell inequality and the presence of nonlocal correlation of two-qubit quantum states.
 
\textbf{Measurement-induced nonlocality:}
Recently, a new measure of quantum correlation has been introduced from the perspective of measurements called measurement-induced nonlocality (MIN). Originally, it was defined as the maximal nonlocal effect due to locally invariant projective measurement via Hilbert-Schmidt norm. Mathematically, the definition of MIN is given as \cite{Luo2011}
\begin{equation}
\mathcal{N}_{2}\mathcal{(\rho)}=~^{\text{max}}_{\Pi ^{a}}\| \rho - \Pi ^{a}(\rho )\| ^{2},
\end{equation}
where $\|\mathcal{O}\| ^{2}=\text{Tr}(\mathcal{O}\mathcal{O}^{\dagger})$ is Hilbert-Schmidt norm of operator $\mathcal{O}$ and the maximum is taken over the locally invariant projective measurements on subsystem $a$ which does not change the state.  The post-measurement is defined as $\Pi^{a}(\rho) = \sum _{k} (\Pi ^{a}_{k} \otimes   \mathds{1} ^{b}) \rho (\Pi ^{a}_{k} \otimes    \mathds{1}^{b} )$, with $\Pi ^{a}= \{\Pi ^{a}_{k}\}= \{|k\rangle \langle k|\}$ being the projective measurements on the subsystem $a$, which do not change the marginal state $\rho^{a}$ locally i.e., $\Pi ^{a}(\rho^{a})=\rho ^{a}$. 

In fact the MIN has a closed formula for any $2 \times n$ dimensional system and is given as 
\begin{equation}
\mathcal{N}_{2}\mathcal{(\rho)} =
\begin{cases}
\text{Tr}(RR^t)-\frac{1}{\| \textbf{x}\| ^2}\textbf{x}^tRR^t\textbf{x}& 
\text{if} \quad \textbf{x}\neq 0,\\
\text{Tr}(RR^t)- R_3&  \text{if} \quad \textbf{x}=0,
\end{cases}
\label{HSMIN}
\end{equation}
where $R_3$ is the least eigenvalue of matrix $RR^t$, the superscript $t$ stands for the transpose of a matrix.
\begin{figure*}[t]
	\centering
	\includegraphics[width=0.9\textwidth]{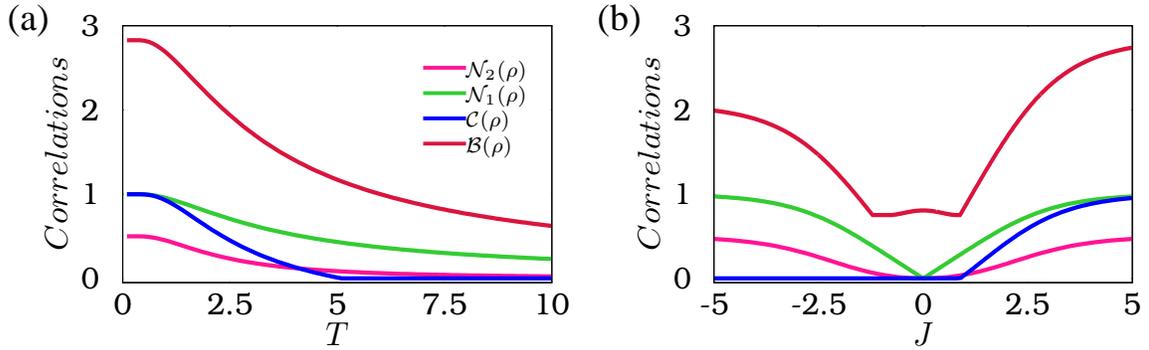}
	\caption{(a) Quantum correlations as a function of temperature $T$ for the state of Heisenberg XY model in the absence of $\cal{PT}$-symmetric operation for the values of $f=1$, $\gamma=0.05$, $B=1.5$, and $J=4.5$. (b)  Quantum correlations versus $J$ at $T=1$. The other parameters are the same as considered in the figure (a). MIN, Trace MIN, Concurrence, and Bell inequality are shown in magenta, green, blue, and red curves. }
	\label{coralp0}
\end{figure*}

\textbf{Trace distance-based MIN}:
The Hilbert-Schmidt norm based measures are not a faithful measure of quantum correlations due to the local ancilla problem, which is indicated by Piani \cite{Piani2012}.  A  natural way to circumvent this issue is to redefine MIN in terms of contractive distance measure. Another alternate form of MIN is based on trace distance ($l_1$ norm) \cite{HFan2015},  namely, trace MIN (T-MIN)  which resolves the local ancilla problem  \cite{HFan2015}. 
The trace distance-based MIN is defined as  \cite{HFan2015}
\begin{equation}
\mathcal{N}_{1}\mathcal{(\rho)}:= ~^{\text{max}}_{\Pi^a}\Vert\rho-\Pi^a(\rho)\Vert_1,
\end{equation} 
where $\Vert \mathcal{O} \Vert_1 = \text{Tr}\sqrt{\mathcal{O}^{\dagger}\mathcal{O}}$ is the trace norm of operator $\mathcal{O}$. Here also, the maximum is taken over all locally invariant projective measurements. For the considered system, the closed formula of trace MIN $\mathcal{N}_{1}\mathcal{(\rho)}$ is given as 
\begin{equation}
\mathcal{N}_{1}\mathcal{(\rho)}=
\begin{cases}
\frac{\sqrt{\chi_+}~+~\sqrt{\chi_-}}{2 \Vert \textbf{x} \Vert_1} & 
\text{if} \quad \textbf{x}\neq 0,\\
\text{max} \lbrace \vert c_1\vert,\vert c_2\vert,\vert c_3\vert\rbrace &  \text{if} \quad \textbf{x}=0,
\end{cases}
\label{TMIN}
\end{equation}
where $\chi_\pm~=~ \alpha \pm 2 \sqrt{\tilde{\beta}}~\Vert \textbf{x} \Vert_1$, $\alpha =\Vert \textbf{c} \Vert^2_1 ~\Vert \textbf{x} \Vert^2_1-\sum_i c^2_i x^2_i$, $\tilde{\beta}=\sum_{\langle ijk \rangle} x^2_ic^2_jc^2_k$, and $c_{i}=Tr(\rho(\sigma_i \otimes \sigma_i))$. Here $\vert c_i \vert $'s are the absolute values of $c_i$ and the summation runs over cyclic permutation of 
$\lbrace 1,2,3 \rbrace$.
\section{The model}
In general, thermal fluctuation degrades the quantum property of the system. To study the role of thermal effects on quantum correlations and teleportation technique, we consider a pair of spin-1/2 particles coupled by Heisenberg interaction. The Hamiltonian of the XY Heisenberg spin model is given as 
\begin{align}
\mathcal{H}=\frac{1}{2}\left[J((1+\gamma)\sigma^x_1\sigma^x_2+(1-\gamma)\sigma^y_1\sigma^y_2)+B(\sigma^z_1+\sigma^z_2)\right]
\end{align}
where $J$ is exchange coupling between the spins, $\gamma$ represents the anisotropy in $xy-$plane, $\sigma_i^\delta~ (i=1,2;~\delta =x,y,z)$ are the usual Pauli's spin matrices and $B$ represents the external magnetic field along the z direction. 
\begin{figure*}[t]
	\centering
	\includegraphics[width=0.9\textwidth]{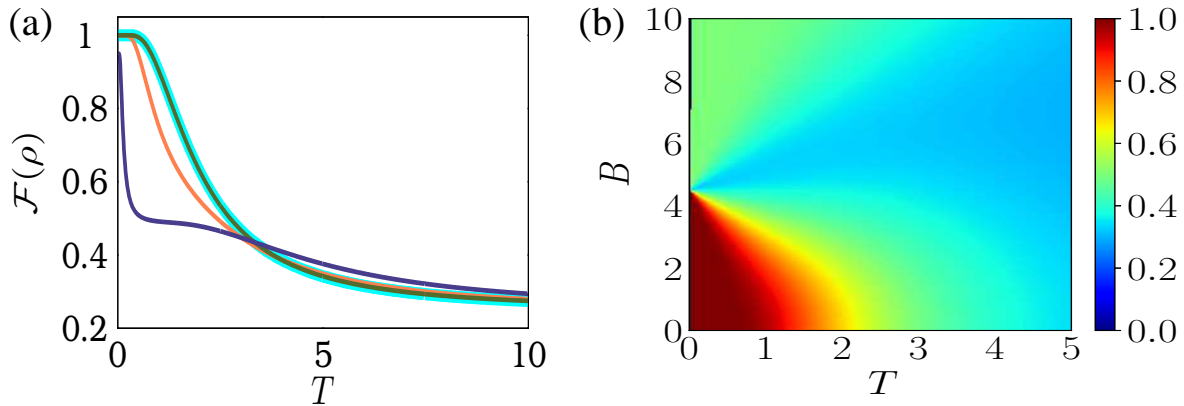}
	\caption{(a) Fidelity of teleportation versus $T$ for different values of $\gamma=-0.01$ (cyan curve),$-0.5$ (orange curve), $0.05$ (green curve), $1.0$ (blue curve) and for the magnetic field $B=1.5$ in the absence of non-unitary time evolution (i.e. $\phi=0$, $t=0$). (b) Thermal teleportation as a function of temperature $T$ and the magnetic field $B$ is plotted in color code. Other parameters are $f=1$, $\gamma=0.05$, and $J=4.5$.}
	\label{fidelalp0}
\end{figure*}
In the standard two-qubit computational basis $\{ |00\rangle, |01\rangle, |10\rangle, |11\rangle \}$, the eigenvalues and the corresponding eigenvectors of the Hamiltonian are computed as 
\begin{align}
   E_{1,2} =\pm J, ~~~ |E_{1,2}\rangle=\frac{1}{\sqrt{2}}\left(|01\rangle \pm |10\rangle \right), \\
     E_{3,4} =\pm \sqrt{\eta},  ~~~ |E_{3,4}\rangle= N_{\pm}\left[(B\pm \sqrt{\eta})|00\rangle + J\gamma |11\rangle \right],
\end{align}
where $\eta = B^2+(J\gamma)^2$ and $N_{\pm}=[(B \pm \sqrt{\eta})^2  +(J \gamma)^2]^{-1/2}$ are the normalization constants. At thermal equilibrium, the thermal density operator is defined as
\begin{align}
    \rho(T)=\frac{1}{\mathcal{Z}}\exp{\left(-\beta\mathcal{H}\right)}=\frac{1}{\mathcal{Z}}\sum_{i=1}^4 \varepsilon_{i} \vert E_{i}\rangle \langle E_{i}\vert
    \label{partition1}
\end{align}
where ${\mathcal{Z}} = \text{Tr}(\exp\left(-\beta \mathcal{H}\right))$ being the canonical ensemble partition function of the system, $\varepsilon_{i}/\mathcal{Z}$ are the eigenvalues of $\rho(T)$ and the inverse thermodynamic temperature $\beta = 1/(k_BT)$ wherein $k_B$ is the Boltzmann's constant which is considered as unity in the following for simplicity. The thermal density matrix $\rho(T)$ can be written as
\begin{align}
\rho(T) = \frac{1}{\mathcal{Z}}\begin{pmatrix}
 \mu_- & 0 & 0 & \nu \\
 0 & \kappa & \omega & 0 \\
 0 & \omega  & \kappa & 0 \\
\nu & 0 & 0 &  \mu_+
\end{pmatrix}.
\label{thermal}
\end{align}
where the matrix elements are $\mu_{\pm}=\text{cosh}(\beta\sqrt{\eta})\pm\frac{B}{\sqrt{\eta}}\text{sinh}(\beta\sqrt{\eta})$, $\kappa=\text{cosh}(J \beta)$, $\omega=-\text{sinh}(J \beta)$, $\nu=-\frac{J\gamma}{\sqrt{\eta}}\text{sinh}(\beta\sqrt{\eta})$ and the partition function ${\mathcal{Z}}=2(\text{cosh}(\beta\sqrt{\eta})+\text{cosh}(J \beta))$.
\section{Results and Discussions}
\subsection{Thermal Quantum Correlations and Teleportation}
In the following, we study the behaviors of thermal quantum correlation of the Heisenberg spin model in terms of entanglement, Bell nonlocality, and  measurement-induced nonlocality (MIN). Fig. 1(a) shows the above-mentioned four correlation measures as a function of the equilibrium temperature $T$. At $T=0$, all the correlation measures are maximum and the state is maximally entangled, which is understood from the spectra of the Hamiltonian. As the temperature increases, all the correlation measures decrease gradually which is mainly due to the influence of thermal fluctuations  in the system. At a sufficiently high temperature $(T=5)$, $\mathcal{C}(\rho)$ only vanishes, and  the other measures (MIN) have nonzero values. To be specific, we observe that at temperature $T\approx 2$, the entanglement  $\mathcal{C}(\rho)$ is non-zero, and $\mathcal{B}(\rho)$ is less than $2$ which indicates that the quantum state has no nonclassical correlation in terms of the Bell nonlocality. Similarly, at $T\approx 5$, the entanglement is zero. On the other hand, both the MIN and trace MIN decrease monotonically with temperature and show more robustness against thermal agitation compared to other measures. 

 In order to understand the role of coupling strength on the correlation between the spins, we have plotted the different correlation measures as a function of $J$  in Fig. 1(b).  From this, we observe that the entanglement is zero in the ferromagnetic phase and nonzero in the anti-ferromagnetic phase. Meanwhile, the thermal state has zero entanglements in the region of $0\leq J\leq0.9$ and for all the negative values of J. The maximal Bell function having a value less than 2 also ensures the absence of nonlocal attributes in the ferromagnetic phase.  Nevertheless, MINs capture nonzero correlation between the spins even in the ferromagnetic region ($J>0$), implying that the measures may signal phase transition between the ferro- and antiferro-magnetic phases. Similar results were observed in the antiferro-magnetic phases. On further increasing the value of $J>0.9$, one can reach the maximum value of entanglement and Bell's inequality which implies that the states are entangled in the anti-ferromagnetic phase. In addition, we note that the anisotropy parameter $\gamma$ also strengthens the quantum correlations. 
 

Next, we study the teleportation of any arbitrary state $\rho_{in}=|\Upsilon \rangle \langle \Upsilon|$ in the Heisenberg XY spin system through Pauli's channels. 
\begin{figure*}[t]
	\centering
	\includegraphics[width=0.9\textwidth]{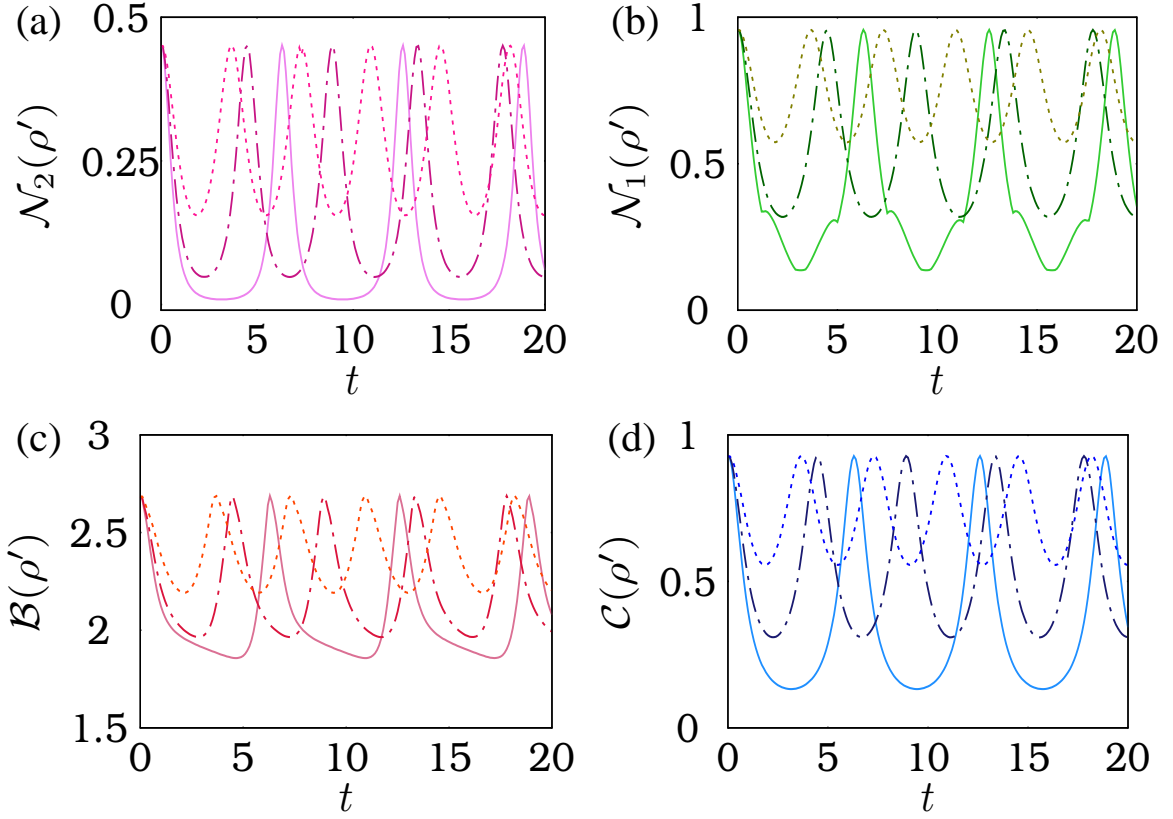}
	\caption{ Quantum correlations with respect to time $t$ for the state of Heisenberg XY model in the presence of $\cal{PT}$-symmetric operation for the values of $T=1$, $f=1$, $\gamma=0.05$, $B=1.5$, and $J=4.5$. (a) MIN and (b) TMIN, (c) Bell inequality, and (d) Concurrence versus time. Solid line, dash-dotted line, and dashed line are plotted for the values of $\phi=\frac{\pi}{3}$, $\phi=\frac{\pi}{4}$, and $\phi=\frac{\pi}{6}$, respectively.}
	\label{coralpnon0}
\end{figure*}
To characterize the efficiency of the teleportation, we employ the fidelity \cite{fidelformula1, fidelformula2, fidelformula3} between $\rho_{out}$ and $\rho_{in}$ defined by
\begin{equation}
\mathcal{F}(\rho)=\left\{\text{Tr}\left(\sqrt{(\rho_{in})^{1/2}\rho_{out}(\rho_{in})^{1/2}}\right)\right\}^{2}
\end{equation}
\par The output state is then given by
\begin{equation}
\rho_{out}=\sum_{mn} p_{mn}(\sigma_{m} \otimes \sigma_{n}) \rho_{in}(\sigma_{m} \otimes \sigma_{n})
\end{equation}
where $\sigma_{m,n} (m,n=0,1,2,3)$ signify the unit matrix and the components of the spin vectors $\vec{\sigma}$, respectively, $p_{mn}=tr[\mathcal{E}^{m} \rho(T)]tr[\mathcal{E}^{n} \rho(T)]$, and $\sum_{mn}p_{mn}=1$. Here $\mathcal{E}^{0}=|\Psi^{-}\rangle \langle \Psi^{-}|$,
$\mathcal{E}^{1}=|\Phi^{-}\rangle \langle \Phi^{-}|$,
$\mathcal{E}^{2}=|\Phi^{+}\rangle \langle \Phi^{+}|$, $\mathcal{E}^{3}=|\Psi^{+}\rangle \langle \Psi^{+}|$, in which $|\Psi^{\pm}\rangle=(1/ \sqrt{2}) (|01\rangle \pm |10\rangle)$,
$|\Phi^{\pm}\rangle=(1/ \sqrt{2}) (|00\rangle \pm |11\rangle)$.

In order to evaluate the above, we consider a pure and maximally  entangled state  $|\Upsilon \rangle=(|00\rangle +|11\rangle)/ \sqrt{2}$ as an input for this process. The fidelity of teleportation for the above considered initial state and output state is displayed in Fig. 2 as a function of temperature and magnetic field. The fidelity of thermal teleportation is maximum at near $T=0$. As the temperature increases, the fidelity decreases monotonically and reaches a non-zero steady state value even at high temperature as shown in Fig. 2(a). The functional behaviors of the teleportation fidelity are exactly the same for the values $ \gamma=-0.01$, and $\gamma=0.05$ and are different for the values $\gamma=-0.5$, and $\gamma=1.0$. More specifically, the maximum value of teleportation fidelity decreases well for the value of $\gamma=1.0$ as compared to the other parametric values of anisotropy. The most interesting fact that we observed here is that entanglement is the ultimate resource for teleportation. Further, Guo-Feng Zhang has observed that in the Heisenberg spin model teleportation is not possible  where the entanglement is zero \cite{DZ}. Interestingly, we found that the fidelity is non-zero even in the absence of entanglement. The next question arises, how does the maximum fidelity of thermal teleportation gets affected when the entangled states are nonlocal in the presence of external magnetic field $B$? To elucidate this, the fidelity of thermal teleportation is presented in Fig. 2(b) as a function of $B$ and $T$. In Fig. 2(b) we have red colour regions where the fidelity is maximum and dark blue colour regions where the fidelity is minimum. For any value of $B$, one can achieve thermal teleportation through the considered channels with nonzero fidelity.
 \begin{figure*}[t]
 	\centering
 	\includegraphics[width=1.0\textwidth]{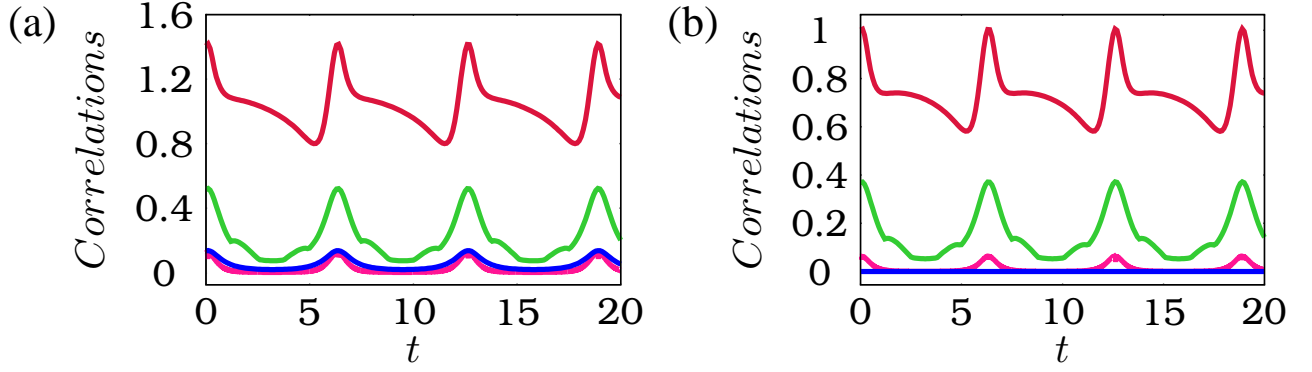}
 	\caption{ Quantum correlations with respect to time $t$ for the state of Heisenberg XY model in the presence of $\cal{PT}$-symmetric operation for $\phi=\frac{\pi}{3}$. (a) at $T=4$ and (b) at $T=6$. MIN, Trace MIN, Concurrence, and Bell inequality are shown in Magenta, green, Blue, and Red curves. Other parametric values are the same as considered in Fig. 3.}
 	\label{corallT4}
 \end{figure*}
\section{Role of $\cal{PT}$-Symmetry}
In the previous section, we have observed that the thermal fluctuation degrades the quantum correlation and efficiency of the teleportation protocol. To minimize the thermal effects on measures, we evolve the thermal state $\rho(T)$ under non-unitary operation. To investigate the dynamical properties of different quantum correlation measures under non-unitary evolution, we consider the $\cal{PT}$-symmetric non-Hermitian Hamiltonian as
\begin{align}
H=\begin{pmatrix}
i f \sin \phi & f\\
f & -i f \sin \phi\\
\end{pmatrix}
\end{align}
where $f$ is a real number with -$\frac{\pi}{2}\leq \phi \leq\frac{\pi}{2}$. Note that, $\phi=\frac{\pi}{2}$ is the $\cal{PT}$-symmetry breaking point. When $\phi=0$, the Hamiltonian $H$ is Hermitian and the $\cal{PT}$-symmetric operation turns into local unitary operation. Based on the non-Hermitian quantum theory, when -$\frac{\pi}{2}<\phi<\frac{\pi}{2}$, the time-evolution operator $U(t)=e^{-iHt}$ has the following form
\begin{align}
U(t)=\frac{1}{\cos\phi}
\begin{pmatrix}
\cos (\psi-\phi) &-i \sin \psi \\
-i \sin \psi & \cos (\psi+\phi)\\
\end{pmatrix}
\label{ut}
\end{align}
with $\psi=f t \cos \phi$. Here, the above considered non-Hermitian system acts as the local environment for the subsystem $a$, and the global state $\rho(T)$ evolves under the nonunitary evolution. 

Using the time evolution operator given in Eq. (\ref{ut}), the final state of the system can be obtained by 
\begin{equation}
\rho^{'}(t)=\frac{(U(t) \otimes \mathds{1})\rho(T)(U(t)^\dagger \otimes \mathds{1})}{Tr[(U(t) \otimes \mathds{1})\rho(T)(U(t)^\dagger \otimes \mathds{1})]}
\label{rhoprime}
\end{equation}
where $\mathds{1}$ is an identity matrix acting on the subsystem $b$ and $\rho(T)$ is as given in Eq. (\ref{thermal}).
\begin{figure*}[t]
	\centering
	\includegraphics[width=0.9\textwidth]{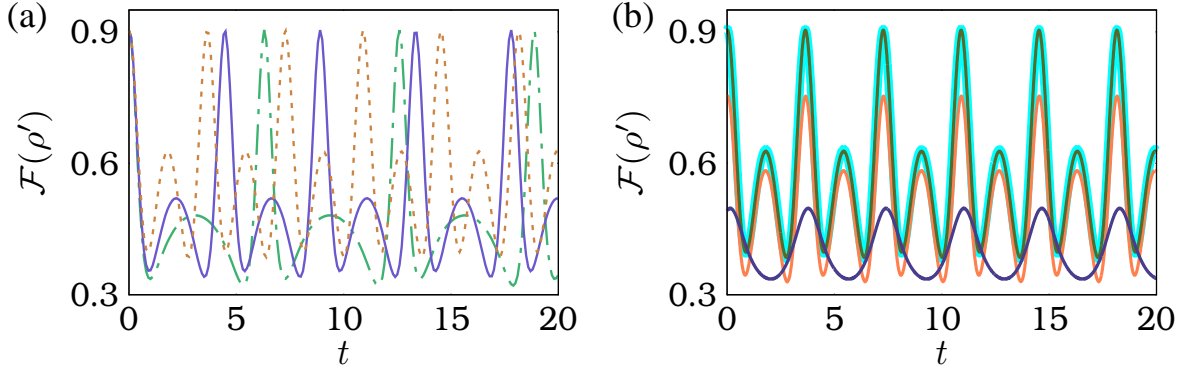}
	\caption{(a) Teleportation fidelity $\cal{F}(\rho^\prime)$ versus time $t$ for the state of Heisenberg XY model under $\cal{PT}$-symmetric operations for various values of $\phi$, $\phi=\frac{\pi}{3}$ (dash-dotted line), $\phi=\frac{\pi}{4}$ (solid line), and $\phi=\frac{\pi}{6}$ (dashed line). (b) $\cal{F}(\rho^\prime)$ versus $t$ for different values of $\gamma$, $\gamma=-0.01$ (cyan curve),$-0.5$ (orange curve), $0.05$ (green curve), $1.0$ (blue curve). By fixing the temperature at $T=1$ and other values as $\phi=\frac{\pi}{6}$, $f=1$, $B=1.5$, and $J=4.5$, we have plotted the figures.}
	\label{fidelalpnon0}
\end{figure*}
 The elements of the matrix given in Eq. (\ref{rhoprime}) are
\begin{eqnarray}
\rho^{'}_{11}&=&\sec (\phi ) \left(\mu_{-} \cos ^2(\phi -\psi )+\kappa \sin ^2(\psi )\right), \nonumber\\ \rho^{'}_{13}&=&i \sec (\phi ) \sin (\psi ) (\mu_{-} \cos (\phi -\psi )-k \cos (\phi +\psi )),\nonumber \\  
\rho^{'}_{33}&=&\sec (\phi ) \left(\mu_{-} \sin ^2(\psi )+\kappa \cos ^2(\phi +\psi )\right), \nonumber \\ 
\rho^{'}_{22}&=&\sec (\phi ) \left(\kappa \cos ^2(\phi -\psi )+\mu_{+} \sin ^2(\psi )\right),\nonumber \\ 
\rho^{'}_{24}&=&i \sec (\phi ) \sin (\psi ) (\kappa \cos (\phi -\psi )-\mu_{+} \cos (\phi +\psi )), \nonumber\\
\rho^{'}_{44}&=&\sec (\phi ) \left(\kappa \sin ^2(\psi )+\mu_{+} \cos ^2(\phi +\psi )\right),\nonumber \\  
\rho^{'}_{12}&=&-i \sin ^2(\psi ) (\omega-\nu) (\tan (\phi )+\cot (\psi )),\nonumber \\ 
 \rho^{'}_{14}&=&\sec (\phi ) \left(\omega \sin ^2(\psi )+\nu \cos (\phi -\psi ) \cos (\phi +\psi )\right),\nonumber \\
 \rho^{'}_{23}&=&\sec (\phi ) \left(\omega \cos (\phi -\psi ) \cos (\phi +\psi )+\nu \sin ^2(\psi )\right), \nonumber \\ 
 \rho^{'}_{34}&=&i \sec (\phi ) \sin (\psi ) (\omega-\nu) \cos (\phi +\psi ),
 \label{ptrhoel}
 \end{eqnarray}
  with a common denominator factor $M_1=\sec (\phi ) (\mu_{-}+2 \kappa+\mu_{+})-\tan (\phi ) ((\mu_{-}+\kappa) \sin (\phi -2 \psi )+(\kappa+\mu_{+}) \sin (\phi +2 \psi )$ for all the matrix elements given above in Eq. (\ref{ptrhoel}). We also note that the above matrix elements satisfy the relation $\rho^{'}_{ij}=\rho^{'*}_{ji}$.
\subsection{Correlations and teleportation under $\cal{PT}$-symmetric operation}
To protect the resources from the thermal effects, we employ the $\cal{PT}$-symmetry as a tool. The quantum correlation quantifiers such as MIN, T-MIN, Bell nonlocality and entanglement (quantified by concurrence) for the considered model after the $\cal{PT}$-symmetric operation are portrayed in Figs. 3(a)-3(d) for different values of $\phi$ with the same parametric values ($J$, $B$ and $\gamma$) as considered in Fig. 1(a) and 2(a). As seen in Fig. 3, at $t=0$, the correlation measures are maximum. For a fixed temperature $T=1$,  we observe that the evolution of correlation measures under the $\cal{PT}$-symmetric operation with respect to time $t$ are in periodic evolution. The frequency of oscillations for all the functions increases monotonically for different strengths of the non-Hermiticity parameter $\phi=\frac{\pi}{3},\frac{\pi}{4},\text{and} ~\frac{\pi}{6}$. However, the maximum value of the amplitude is the same for all the values of $\phi$. 

Further, the violation of Bell's inequality during the time evolution now indicates the nonlocality of the considered system. Note that the Bell's inequality ceases to violate with respect to temperature before the inclusion of $\cal{PT}$-symmetric operation. More interestingly, the value of $\mathcal{C}(\rho^\prime)$ corresponding to the temperature which is frozen to zero as shown in the dark blue curves in Fig. 1(a) in the absence of $\cal{PT}$-symmetric operation now turns to oscillate periodically under the non-unitary time evolution as depicted in Fig. 3(d). From the results, we observe that the change of correlation measures is periodic with time $t$ and it persists for a long enough time. The recovery of Bell's nonlocality and concurrence is possible for a certain range of time and thus we get the maximum value of Bell nonlocality and concurrence by choosing the suitable value of $t$ under $\cal{PT}$-symmetric operation. 

A similar analysis is also done for higher temperatures such as $T=4$ and $T=6$ and the correlation measures are plotted in Fig. 4.  We observe that the quantum correlation measures exhibit similar  dynamics. While comparing the correlation measures in the absence of $\cal{PT}$-symmetric operation, the measures are now greatly enhanced due to the non-unitary evolution. Here again, the entanglement and Bell nonlocality fail to quantify the quantum correlation at high temperatures. On the other hand, the $\cal{PT}$-symmetry operation strengthens the quantum correlation measures. It is worth mentioning that the noise causes sudden death of entanglement \cite{pra, suddendeath}. On the other hand, the dynamics of entanglement is protected under non-unitary ($\cal{PT}$-symmetric) operation. Notably, similar results were observed by performing a local non-Hermitian on two qubit entangled system \cite{enhanceprotect}.

Next, we investigate the role of $\cal{PT}$-symmetry in the teleportation technique.  Fig. 5(a) shows the time evolution of fidelity of teleportation $\mathcal{F}(\rho^\prime)$ for different strengths of non-Hermiticity. The fidelity of teleportation, which shows decaying dynamics in the absence of $\cal{PT}$-symmetry now changes into periodic oscillations. The maximum value of fidelity is the same as before and after the operation of $\cal{PT}$-symmetry. However, as the degree of non-Hermiticy decreases, the frequency of oscillations increases with the same value of maximum amplitude.  Fig. 5(b) is plotted for $\mathcal{F}(\rho^\prime)$ with different values of $\gamma$. The maximum value of fidelity increases with the lower value of $\gamma$ mainly, say $\gamma=-0.01$, as shown in the cyan colour curve in Fig. 5(b). The oscillation pattern is the same in terms of amplitude and frequency as observed for the value of $\gamma=0.05$, as shown in green colour curves in Fig. 5(b). Conversely, the maximum fidelity decreases with the values of $\gamma=-0.5$ and $\gamma=1.0$ as shown in orange and blue colour curves in Fig. 5(b). In Fig. 6, the fidelity under $\cal{PT}$-symmetric operation is  observed for higher temperatures $T=4$ and $T=6$, which exhibits the same behaviour as shown in the case of low temperature $T=1$. From this investigation, we have found that performing non-unitary time evolution on thermally entangled states is useful for the revival of the teleportation process with maximum fidelity at certain values of $t$.

\begin{figure*}[t]
	\centering
	\includegraphics[width=0.9\textwidth]{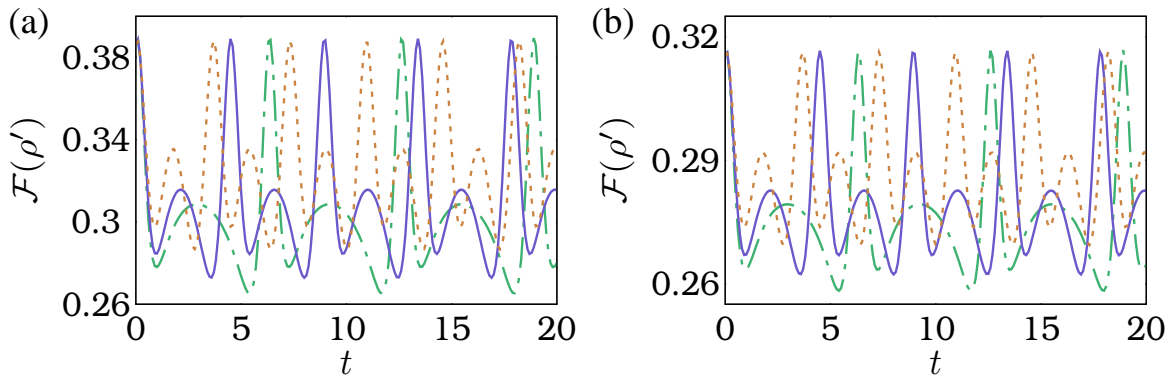}
	\caption{Teleportation fidelity $\cal{F(\rho^\prime)}$ for the state of Heisenberg XY model under $\cal{PT}$-symmetric operations with respect to time $t$ for various values of $\phi$, $\phi=\frac{\pi}{3}$ (dash-dotted line), $\phi=\frac{\pi}{4}$ (solid line), and $\phi=\frac{\pi}{6}$ (dashed line). (a) for $T=4$, and (b) for $T=6$. The figures are plotted for $f=1$, $B=1.5$, $\gamma=0.05$, and $J=4.5$.}
	\label{fidelalpnon0}
\end{figure*}
\section{Conclusions}
In this paper, we have examined the quantum correlation measures and teleportation techniques in the Heisenberg spin model with and without $\cal{PT}$-symmetry. We have demonstrated the ability of different criteria to capture the nonlocal attributes of the Heisenberg spin model. We captured the entire spectrum of nonlocal aspects of the system in terms of measurement-induced nonlocality even in the absence of entanglement. This result also suggests that the absence of entanglement does not indicate the absence of nonlocal properties of the quantum system. Moreover, we have implemented the teleportation protocol in the spin system successfully even in the absence of entanglement. 

We have also studied the behavior of quantum correlation and quantum teleportation under non-unitary evolution. In the light of our results, we show that $\cal{PT}$-symmetry greatly enhances the quantum correlations and efficiency of teleportation techniques which have potential applications in quantum communication. Further, our exploitation highlights that efficient information processing based on the measurement-induced nonlocality offer more resistance to non-unitary time evolution and are completely distinguishable from that of entanglement.

\section*{Acknowledgements}
J.R.P. thanks the Department of Science and Technology, Government of India, for providing an INSPIRE Fellowship No. DST/INSPIRE Fellowship/2017/IF170539. R.M. is grateful for the CTU Global Postdoc Fellowship Program and the financial support from MŠMT RVO 14000. The work of V.K.C. forms part of the research projects sponsored by SERB-DST-MATRICS Grant No. MTR/2018/000676 and CSIR Project Grant No. 03(1444)/18/EMR-II. J.R.P., and V.K.C. also wish to thank DST, New Delhi for computational facilities under the DST-FIST programme (SR/FST/PS- 1/2020/135) to the Department of Physics. The work of M.L. is supported by a DST-SERB National Science Chair position (NSC/2020/000029).

\end{document}